\documentstyle[12pt,prd,aps,epsfig,preprint]{revtex}
\begin{document}
\draft
\title{Isovector Collective Response Function of Nuclear
       Matter at Finite Temperature}
\author{
Kutsal Bozkurt   }
\address{Physics Department, Middle East Technical University,
06531 Ankara, Turkey \\
kutsalb@metu.edu.tr }
\date{\today}
\maketitle
\begin{abstract}
We study isovector collective excitations in nuclear matter by
employing the linearized Landau-Vlasov equation with and without a
non-Markovian binary collision term at finite temperature. We
calculate the giant dipole resonance (GDR) strength function for
finite nuclei using Steinwedel-Jensen model and also by
Thomas-Fermi approximation, and we compare them for $^{120}Sn$ and
$^{208}Pb$ with experimental results.
\end{abstract}

~~~~~~~~

\narrowtext
\newpage

Giant resonances, in particular giant dipole resonance (GDR),
built on highly excited nuclear states  have been the subject of
many experimental and theoretical studies in recent
years\cite{R1,R2,R3}. The damping properties and the excitation
energy dependence of GDR width are still among the open problems
in nuclear collective dynamics. There are essentially two
different theoretical approaches to this problem. The first one
explains the temperature  dependence of the width by the coherent
mechanism due to adiabatic coupling of the collective state with
thermal surface deformations \cite{R4}. In the second approach,
referred to as collisional damping, the coupling with incoherent
two particle-two hole states plus a Landau damping form the
mechanism for temperature dependence
\cite{R5,R6,R7,R8,R9,R10,R11}.

Investigation of the GDR strength function has been carried out in
the mean-field approximation without the collisional damping in
\cite{R10}. In a previous work, we investigated the GDR strength
function in infinite nuclear matter in quantal framework by
employing the linearized version of the extended time dependent
Hartree-Fock theory (TDHF) with a non-Markovian binary collision
term \cite{R12}. We then applied our results to finite nuclei
using the Steinwedel-Jensen model. However, this model is not very
reliable for the treatment of collective dipole oscillations of
finite nuclei since in this model it is assumed that the nuclear
surface remains constant and the density oscillations obeys a wave
equation with the boundary condition that the radial velocity
vanishes on the spherical surface with radius $R=R_{0}A^{1/3}$. On
the other hand, semi-classical approaches based on Thomas-Fermi
approximation which are easier to handle than quantal calculations
have been very useful to tackle the problems related to finite
nuclei \cite{R13}. Indeed, the long wave length limit of TDHF
equation is the Landau-Vlasov equation which is valid if the
spatial variations are slow. Moreover, if Thomas-Fermi method is
used to determine the nuclear density then the Landau-Vlasov
equation forms a better approximation to quantal theory even if
the spatial variations are not small \cite{R13}.

In this work, we study the GDR strength function of finite nuclei
by employing the Landau-Vlasov equation. We do not consider the
coherent mechanism and investigate the temperature dependence due
to coupling of the collective state with incoherent two
particle-two hole states using a non Markovian collision term in
the linearized Landau-Vlasov equation. We apply our results to
finite nuclei using Steinwedel-Jensen model. We then consider
giant dipole excitations in finite nuclei by expanding our results
using Thomas-Fermi approximation. This way, we not only asses the
effects of the collision term but we also compare the differences
between infinite nuclear matter and finite nuclei results.

The transport equation \cite{R14}, for the particle  phase-space
density $f(\vec{r}, \vec{p}, t)$ with collision term is the
Boltzmann equation
\begin{equation}\label{e6}
\frac{\partial}{\partial t}f(\vec{r}, \vec{p},
t)+\vec{\nabla}_{p}~\epsilon(\vec{r},\vec{p}, t).\vec{\nabla}_{r}
f(\vec{r},\vec{p}, t)- \vec{\nabla}_{r}~\epsilon(\vec{r},\vec{p},
t).\vec{\nabla}_{p}f(\vec{r}, \vec{p}, t)=K(f)~~.
\end{equation}
In the framework of Fermi liquid theory defining the quasiparticle
velocity as $\vec{v} =\vec{\nabla}_{p}~\epsilon(\vec{r},\vec{p},
t)$ and assuming that the potential energy $U$ in the Hartree-Fock
Hamiltonian is local, one immediately obtains the transport
equation with collision term for proton  $(q=p)$ and neutron
$(q=n)$ distribution functions $f_{q}$
\begin{equation}\label{e7}
\frac{\partial}{\partial t}f_{q}-\{h_{q}+V_{q}, f_{q} \}_{\vec{r},
\vec{p}}=K(f_{q})
\end{equation}
or
\begin{equation}\label{e8}
\frac{\partial}{\partial t}f_{q}+\vec{v_{q}}.\vec{\nabla}_{r}
f_{q}-
\vec{\nabla}_{r}~(U_{q}+V_{q}).\vec{\nabla}_{p}f_{q}=K(f_{q})~~,
\end{equation}
where $h=T+U$ is the mean-field Hamiltonian, $T$ is the kinetic
energy, $V$ is the external field, and $K(f)$ is the non-Markovian
collision term. When $f$ and $U$ change by a small amount around
the equilibrium, we then have
\begin{equation}\label{e9}
f(\vec{r},\vec{p}, t)= f_{eq}(\epsilon_{p})+\delta
f(\vec{r},\vec{p}, t)~,~~~U(\vec{r}, t)= U_{0}(\vec{r})+\delta
U(\vec{r},t)
\end{equation}
with
\begin{equation}\label{e10}
f_{eq}(\epsilon_{p})=\frac{1}{(1+e^{\beta(\epsilon_{p}-\mu)})}~,~~~
\delta U(\vec{r}, t) =\left(\frac{\partial U}{\partial
\rho}\right)_{\rho_{0}}\delta\rho(\vec{r},
t)~,~~~\epsilon_{p}=\frac{p^{2}}{2m}~~.
\end{equation}
The equation of motion of the small amplitude vibrations in the
semi-classical limit for infinite nuclear matter is then obtained
as
\begin{equation}\label{e11}
\frac{\partial}{\partial t}\delta f+\vec{v}.\vec{\nabla}_{r}
\delta f- \vec{\nabla}_{r}~[\delta U+2 \delta
V].\vec{\nabla}_{p}f_{eq}(\epsilon_{p})=\delta K
\end{equation}
which is  the well-known the linearized Landau-Vlasov equation
with a collision term. In this equation $\delta f \equiv \delta
f_{n}-\delta f_{p}$~, $\delta U \equiv \delta U_{n}-\delta U_{p}$
and $\delta V \equiv \delta V_{n}-\delta V_{p}$ ($\delta
V_{n,p}=\pm  \delta V$) are differences between the indicated
neutron and proton functions.

The isovector mean field $\delta U(\vec{r}, t)$ can be expressed
as \cite{R7,R11}
\begin{equation}\label{e12}
\delta U(\vec{r}, t)= f_{0}\delta\rho(\vec{r}, t)~~~
\end{equation}
where $f_{0} =F'_{0}(T)/N(T)$ is the quasiparticle zero-order
interaction amplitude, $F'_{0}(T)$ is the isovector Landau
parameter
\begin{equation}\label{e13}
F'_{0}(T)\simeq
F'_{0}(T=0)\left[1-\frac{\pi^2}{12}\left(\frac{T}{\epsilon_{F}}\right)^2\right]~~,
\end{equation}
\begin{equation}\label{e14}
\delta\rho(\vec{r}, t)= \int \frac{g
d\vec{p}}{(2\pi\hbar)^3}\delta f(\vec{r},\vec{p}, t)
\end{equation}
is the density distribuiton function, $g=2$ is the spin degeneracy
factor and
\begin{equation}\label{e15}
N(T)= \int \frac{g d\vec{p}}{(2\pi\hbar)^3}\left(-\frac{\partial
f_{eq}(\epsilon_{p})}{\partial\epsilon_{p}}\right)
\end{equation}
is the thermally averaged density of states. For $T=0$ $N(0)$ is
given as $N(0)=g p_{F}m/2\pi^2\hbar^3$, with  $p_{F}$ Fermi
momentum.

Now, we present the details of the calculation of GDR response
function for infinite and finite nuclear matter by using the
linearized Landau-Vlasov equation without and with a non-Markovian
collision term. The solution of Eq. (11) can be found in form of a
plane wave for infinite nuclear matter
\begin{equation}\label{e16}
\delta f(\vec{r},\vec{p}, t)=\delta
f_{k,\omega}(\vec{p})~e^{i[\vec{k}.\vec{r}-(\omega+i\eta)t]}~~,
\end{equation}
where $\eta$ is the vanishingly small positive number
corresponding to an adiabatic switching of the field at time
$t=-\infty$. From the collisionless Landau-Vlasov equation, we
have
\begin{equation}\label{e17}
\delta f(\vec{r},\vec{p},
t)+\frac{\vec{k}.\vec{v}}{\omega+i\eta-\vec{k}.\vec{v}}\frac{\partial
f_{eq}(\epsilon_{p})}{\partial\epsilon_{p}}[\delta U+2\delta
V]=0~~,
\end{equation}
and then integrating $\int \frac{g d\vec{p}}{(2\pi\hbar)^3}$ with
weight 1, we obtain
\begin{equation}\label{e18}
\delta\rho+[f_{0}\delta\rho+2\delta V]\chi^{(1)}(\vec{k},
\omega)=0~~,
\end{equation}
where
\begin{equation}\label{e19}
\chi^{(1)}(\vec{k}, \omega)=\int \frac{g
d\vec{p}}{(2\pi\hbar)^3}\frac{\vec{k}.\vec{v}}{\omega+i\eta-\vec{k}.\vec{v}}\frac{\partial
f_{eq}(\epsilon_{p})}{\partial\epsilon_{p}}
\end{equation}
is the unperturbed Lindhard function. We can then write response
of the collisionless system for an external field $\delta V
\propto e^{i[\vec{k}.\vec{r}-(\omega+i\eta)t]}$ as
\begin{equation}\label{e20}
\Pi^{0}(\vec{k}, \omega)=-\frac{\delta \rho}{\delta
V}=\frac{2\chi^{(1)}(\vec{k}, \omega)}{1+f_{0}\chi^{(1)}(\vec{k},
\omega)}~~.
\end{equation}
Performing integrations in Eq.(14), we can  find the real and
imaginary parts of  $\chi^{(1)}(\vec{k}, \omega)$ as (for details
please refer to \cite{R7,R11})
\begin{equation}\label{e21}
Im \chi^{(1)}(\vec{k},
\omega)=-\frac{\pi}{2}N(0)s\left[\frac{m\omega}{k
p_{F}}\right]f_{eq}(s^{2}\bar{\epsilon})
\end{equation}
and
\begin{equation}\label{e22}
Re \chi^{(1)}(\vec{k},
\omega)=\frac{N(0)}{4}\left[\frac{m\omega}{k
p_{F}}\right]f_{eq}(s^{2}\bar{\epsilon})~~.
\end{equation}
Here,
\begin{equation}\label{e23}
\bar{\epsilon}=\frac{5}{3\rho_{eq}}\int \frac{g
d\vec{p}}{(2\pi\hbar)^3}\epsilon_{p}f_{eq}(\epsilon_{p})~~,
\end{equation}
\begin{equation}\label{e24}
\rho_{eq}=\int \frac{g
d\vec{p}}{(2\pi\hbar)^3}f_{eq}(\epsilon_{p})
\end{equation}
are quasiparticle average kinetic energy and density, respectively
and
\begin{equation}\label{e25}
s=\frac{m\omega}{k
p_{F}}\left(\frac{e_{F}}{\bar{\epsilon}}\right)^{1/2}~~.
\end{equation}
The strength distribution function is obtained from the imaginary
part of the  response function \cite{R15}
\begin{equation}\label{e26}
S(\vec{k},w)=-\frac{1}{\pi} Im \Pi^{0}(\vec{k},
\omega)=-\frac{1}{\pi}\frac{2Im \chi^{(1)}(\vec{k},
\omega)}{(1+f_{0}Re \chi^{(1)}(\vec{k}, \omega))^{2}+(f_{0}Im
\chi^{(1)}(\vec{k}, \omega))^{2}}~~.
\end{equation}

The solution of the linearized Landau-Vlasov equation with
collision term for the infinite nuclear matter is given as
\begin{equation}\label{e27}
-i(\omega+i\eta)\delta f+i\vec{k}.\vec{v}\delta
f-i\vec{k}.\vec{v}\frac{\partial
f_{eq}(\epsilon_{p})}{\partial\epsilon_{p}}[f_{0}\delta\rho+2\delta
V]=\delta K~~,
\end{equation}
from which we obtain
\begin{equation}\label{e28}
\delta\rho+[f_{0}\delta\rho+2\delta V]\chi^{(1)}(\vec{k},
\omega)=-[f_{0}\delta\rho+2\delta V]\chi^{(2)}(\vec{k}, \omega)~~.
\end{equation}
The collisional response function $\chi^{(2)}(\vec{k}, \omega)$
can be expressed as\cite{R12}
\begin{equation}\label{e29}
\chi^{(2)}(\vec{k}, \omega)=\frac{1}{(2\pi\hbar)^3}\int
d^3p_1d^3p_2d^3p_3d^3p_4\left(\frac{\Delta
Q}{2}\right)^2\frac{W(12;34)}{\pi}~
\frac{f_1f_2\overline{f}_3\overline{f}_4-
\overline{f}_1\overline{f}_2f_3f_4} {
w-\epsilon_3-\epsilon_4+\epsilon_1+\epsilon_2+i\eta}
\end{equation}
where $f_i=f_{eq}(\epsilon_{i})$,
$\overline{f}_i=1-f_{eq}(\epsilon_{i})$, ~$\Delta
Q=Q_1+Q_2-Q_3-Q_4$ with
$Q_i=1/\left[w-\vec{k}.\vec{v_{i}}\right]$, $\epsilon_{i}=(m/2)
v^{2}_{i}$ and W(12;34) denotes  the basic two-body transition
rate
\begin{equation}\label{e30}
W(12;34)=\frac{\pi}{(2\pi\hbar)^6}|<\frac{\vec{p_{1}}-\vec{p_{2}}}{2}|v|\frac{\vec{p_{3}}-\vec{p_{4}}}{2}>|^{2}
\delta^3(\vec{p}_1+\vec{p}_2-\vec{p}_3-\vec{p}_4)
\end{equation}
which can be expressed in terms of the scattering cross-section as
\begin{equation}\label{e31}
W(12;34)=\frac{1}{(2\pi\hbar)^3}~\frac{4\hbar}{m^2}~
\left(\frac{d\sigma}{d\Omega}\right)_{pn}~
\delta^3(\vec{p}_1+\vec{p}_2-\vec{p}_3-\vec{p}_4)~~.
\end{equation}
Then the retarded response function with collision term is
obtained as
\begin{equation}\label{e32}
\Pi^{coll}(\vec{k}, \omega)=-\frac{\delta \rho}{\delta
V}=\frac{2\chi(\vec{k}, \omega)}{1+f_{0}\chi(\vec{k}, \omega)}
\end{equation}
with $\chi(\vec{k}, \omega)=\chi^{(1)}(\vec{k},
\omega)+\chi^{(2)}(\vec{k}, \omega)$. Thus we can rewrite the
strength distribution for our system with collision term as
\begin{equation}\label{e33}
S(\vec{k},w)=-\frac{1}{\pi} Im \Pi^{coll}(\vec{k},
\omega)=-\frac{1}{\pi}\frac{2Im \chi(\vec{k}, \omega)}{(1+f_{0}Re
\chi(\vec{k}, \omega))^{2}+(f_{0}Im \chi(\vec{k}, \omega))^{2}}~~.
\end{equation}
The strength function satisfies the following energy weighted sum
rule\cite{R10}
\begin{equation}\label{e34}
\int_{0}^{\infty} d\omega \omega S(\vec{k},
\omega)=\frac{k^2}{2m}\rho_{0}
\end{equation}
where $\rho_{0}=0.16 fm^{-3}$ is the saturation density of
infinite nuclear matter.

In the calculations of the collisional response function, we use
conservation laws and symmetry properties. It is possible to
reduce the twelve dimensional integrals to five fold integrals by
the total and relative momentum transformations
($\vec{P}=\vec{p}_1+\vec{p}_2$, $\vec{P'}=\vec{p}_3+\vec{p}_4$,
and relative momenta $\vec{q}=(\vec{p}_1-\vec{p}_2)/2$,
$\vec{q}~^\prime=(\vec{p}_3-\vec{p}_4)/2$) before and after the
collisions. We neglect the real part $Re\chi^{(2)}(\vec{k},
\omega)$ of the function $\chi^{(2)}(\vec{k}, \omega)$ in our
calculations. The collisional response function
$\chi^{(2)}(\vec{k}, \omega)$ has a singular behavior arising from
the pole of the distortion functions, $Q_i=1/\left[
w-\vec{k}.\vec{v_{i}}\right]$. We avoid this singular behavior by
incorporating a pole approximation. In the distortion functions,
we make the replacement $\omega\rightarrow \omega_D-i\Gamma/2$
where $\omega_D$ and $\Gamma$ are determined from
$1+f_{0}\chi^{(1)}(\vec{k}, \omega)=0$ at each temperature that is
considered. So, we evaluate the remaining five dimensional
integrals numerically by employing a fast algorithm. In the
evaluation of momentum integrals we make the replacement
$\left(d\sigma/d\Omega\right)_{pn}\rightarrow\sigma_{pn}/4\pi$
with $\sigma_{pn}=40$ mb, thus neglecting  the angular anisotropy
of the cross section.

In order to apply our results to nuclear dipole vibrations  and
finite nuclei, we work within the framework of Steinwedel and
Jensen model which describes the GDR in heavy nuclei as a volume
polarization mode conserving the total density
$\rho_{0}=\rho_{n}+\rho_{p}$ for infinite nuclear matter
\cite{R16} where neutron and proton oscillate inside a sphere of
radius $R$ as
\begin{equation}\label{e35}
\rho_p(\vec{r},t)-\rho_n(\vec{r},t)\propto\sin
(\vec{k}\cdot\vec{r})e^{iwt}~~.
\end{equation}
According to this model, we choose the wave number of the normal
mode as $k=\pi/2R$. We apply Steinwedel and Jensen model to GDR in
$^{120}Sn$ and $^{208}Pb$, and we take $R=5.6$ fm $k=0.28~fm^{-1}$
for $^{120}Sn$ and $R=6.7$ fm $k=0.23~fm^{-1}$ for $^{208}Pb$
according to $R=1.13A^{1/3}$.

The Landau parameter $F'_{0}(T=0)$ can be expressed as a function
of the symmetry energy coefficient $a_{\tau}$ in the Weizsäcker
mass formula at zero temperature as follows \cite{R17}
\begin{equation}\label{e36}
F'_{0}(T=0)=\frac{3a_{\tau}}{\epsilon_{F}}-1~~.
\end{equation}
For the value of $a_{\tau}= 28~MeV$ we have $F'_{0}(T=0)=1.33$.
The value of the $F'_{0}(T)$ decreases with temperature because of
the decrease of the thermally averaged level density $N(T)$
\cite{R7}.

So far, our GDR calculations have been for  infinite nuclear
matter. In the rest of the paper we will employ  the Thomas-Fermi
approximation (TF) to calculate GDR response function by using the
linearized Landau-Vlasov equation with and without the collision
term at finite temperature for finite nuclei. The Thomas-Fermi
theory, together with its extensions, is the semiclassical
treatment of nuclear dynamics in its independent particle or
Hartree-Fock approximation and can be explained from quite
different points of view \cite{R15,R16}. We evaluate the
expression for the GDR  in the TF approximation, which corresponds
to a semi-classical transport description of the collective
vibrations. The solution of the linearized Landau-Vlasov equation
with and without collision term
\begin{equation}\label{e37}
\frac{\partial}{\partial t}\delta f+\vec{v}.\vec{\nabla}_{r}
\delta f- \vec{\nabla}_{r}~[\delta U+2 \delta
V].\vec{\nabla}_{p}f_{eq}(\epsilon_{p}, r)=\delta K
\end{equation}
for finite nuclei is obtained with the local plane wave ansatz
\begin{equation}\label{e38}
\delta f(\vec{r},\vec{p}, t)=\delta f_{k,\omega}(\vec{r}, \vec{p}
)~e^{i[\vec{k}.\vec{r}-(\omega+i\eta)t]}~~.
\end{equation}
Response of the finite system without and with collision term is
then obtained as
\begin{equation}\label{e39}
\Pi^{0}_{TF}(\vec{k}, \omega)=\frac{2\chi^{(1)}_{TF}(\vec{k},
\omega)}{1+f^{TF}_{0}\chi^{(1)}_{TF}(\vec{k},
\omega)}~~,~~\Pi^{coll}_{TF}(\vec{k},
\omega)=\frac{2\chi_{TF}(\vec{k},
\omega)}{1+f^{TF}_{0}\chi_{TF}(\vec{k}, \omega)}~~,
\end{equation}
where $\chi_{TF}(\vec{k}, \omega)=\chi^{(1)}_{TF}(\vec{k},
\omega)+\chi^{(2)}_{TF}(\vec{k}, \omega)$. The strength
distribution function without and with collision term are
\begin{equation}\label{e40}
S_{TF}(\vec{k},w)=-\frac{1}{\pi} Im \Pi^{0}_{TF}(\vec{k},
\omega)~~,~~S_{TF}(\vec{k},w)=-\frac{1}{\pi} Im
\Pi^{coll}_{TF}(\vec{k}, \omega)
\end{equation}
with
\begin{equation}\label{e41}
\chi^{(i)}_{TF}(\vec{k}, \omega)=\frac{1}{A}\int
d\vec{r}~\rho(r)~\chi^{(i)}(\vec{k}, \omega, r)
\end{equation}
where $i=1,2$~~~,~~~and
\begin{equation}\label{e42}
f^{TF}_{0}=\frac{1}{A}\int d\vec{r}~\rho(r)~f_{0}(r)~~.
\end{equation}
The function $\chi^{(2)}(\vec{k}, \omega, r)$ is obtained by
evaluating the collision term given in Eq. (24) using Thomas-Fermi
approximation. We determine the nuclear density $\rho (r)$ for the
finite nuclear matter in TF approximation using a Wood-Saxon
potential with a depth $V_{0}=44~MeV$, thickness parameter
$t_{p}=0.67~fm$ and sharp radius $R=1.13A^{1/3}$\cite{R16},
\begin{equation}\label{e43}
V(r)=-\frac{V_{0}}{(1+e^{\frac{r-R}{t_{p}}})}~~,~~\rho(r)=\frac{2}{3\pi^2}k_{F}(r)^{3}\Theta(\lambda-V(r))~~,~~k_{F}(r)=\left(\frac{2m}{\hbar^{2}}[V(r_{c})-V(r)]\right)^{1/2}
\end{equation}
where $r_{c}$ is the critical radius for a mass number $A$ is
defined as
\begin{equation}\label{e44}
A=\int_{0}^{r_{c}}d\vec{r} \rho(r)~~.
\end{equation}
Here $\lambda=V(r_{c})$ and, the expression for $A$ can be
numerically integrated for a given $A$ to determine $r_{c}$. For
finite nuclear matter, the interaction amplitude is $f_{0}(r)=
3V^{'}_{0}(r)\rho(r)/(2\epsilon_{F}(r)~N(r, T=0))$ which is
related to the parameters of the simplified Skyrme force as
\begin{equation}\label{e45}
V^{'}_{0}(r)=-\frac{1}{2}t_{0}(x_{0}+\frac{1}{2})-\frac{1}{8}t_{3}\rho(r)~~.
\end{equation}
We use the following parameters: $t_{0}=-983.4~ MeVxfm^{3}$,
$t_{3}=13106~ MeVxfm^{6}$, and $x_{0}=0.48$ \cite{R10}.

We show our results for the GDR strength function with and without
the collision term in Fig. 1 and in Fig. 2 for $^{120}Sn$ and
$^{208}Pb$, respectively, calculated using infinite nuclear matter
formalism within the framework of Steinwedel-Jensen model. In Fig.
3 and in Fig. 4 we show the GDR strength function with and without
collision term for $^{120}Sn$ and $^{208}Pb$, respectively,
calculated employing Thomas-Fermi approximation. In this figures,
we also compare our results with the normalized experimental data
taken from \cite{R1}. The temperature parameter $T$ in the mean
occupation number functions $f(\epsilon, T)$ is related to the
experimental temperature $T^*$ as $T=T^{*}\sqrt{a_{E}/a_{F}}$
where $a_{F}=A\pi^{2}/4\epsilon_{F}$ is the Fermi gas level
density parameter and $a_{E}$ is the energy dependent empirical
level density parameter \cite{R1}.

From these figures, we first note that without the collision term
the position of the peak of the strength function does not change
appreciably with temperature. This behavior is in accordance with
the experimental results \cite{R1,R2}. Since we neglect the real
part of the collisional response, when we include the collisional
term the average position of the peak values of the strength
functions do not change again, but as the result of the collisions
the overall shape of the strength function changes somewhat and
this change becomes more pronounced with increasing temperature.
In the case of infinite nuclear matter this change has the
tendency to improve the agreement with experimental results for
$^{120}Sn$ but this tendency is much less pronounced for
$^{208}Pb$ as it can be seen by comparing Fig. 1 with Fig. 2. On
the other hand, for this case of finite nuclei calculations
employing Thomas-Fermi approximation the change produced by the
addition of the collison term can be clearly noted in Fig. 3 for
$^{120}Sn$ and in Fig. 4 for $^{208}Pb$. Indeed, in both cases,
the overall agreement with the experimental results is much more
improved when the binary collision term is included.

In our work, we obtain a reasonable description of the giant
dipole excitations in $^{120}Sn$ and $^{208}Pb$ using
semi-classical approach with Thomas-Fermi approximation as
compared to infinite nuclear matter formalism, and we demonstrate
the importance of the collision term which improves the agreement
of the calculated strength functions with the experimental
results. We believe that inclusion of the coherent damping
mechanism into our formalism will extend our description further.

\newpage
\begin{figure}
\vspace*{0.0cm}\hspace{-3.8cm}
\epsfig{figure=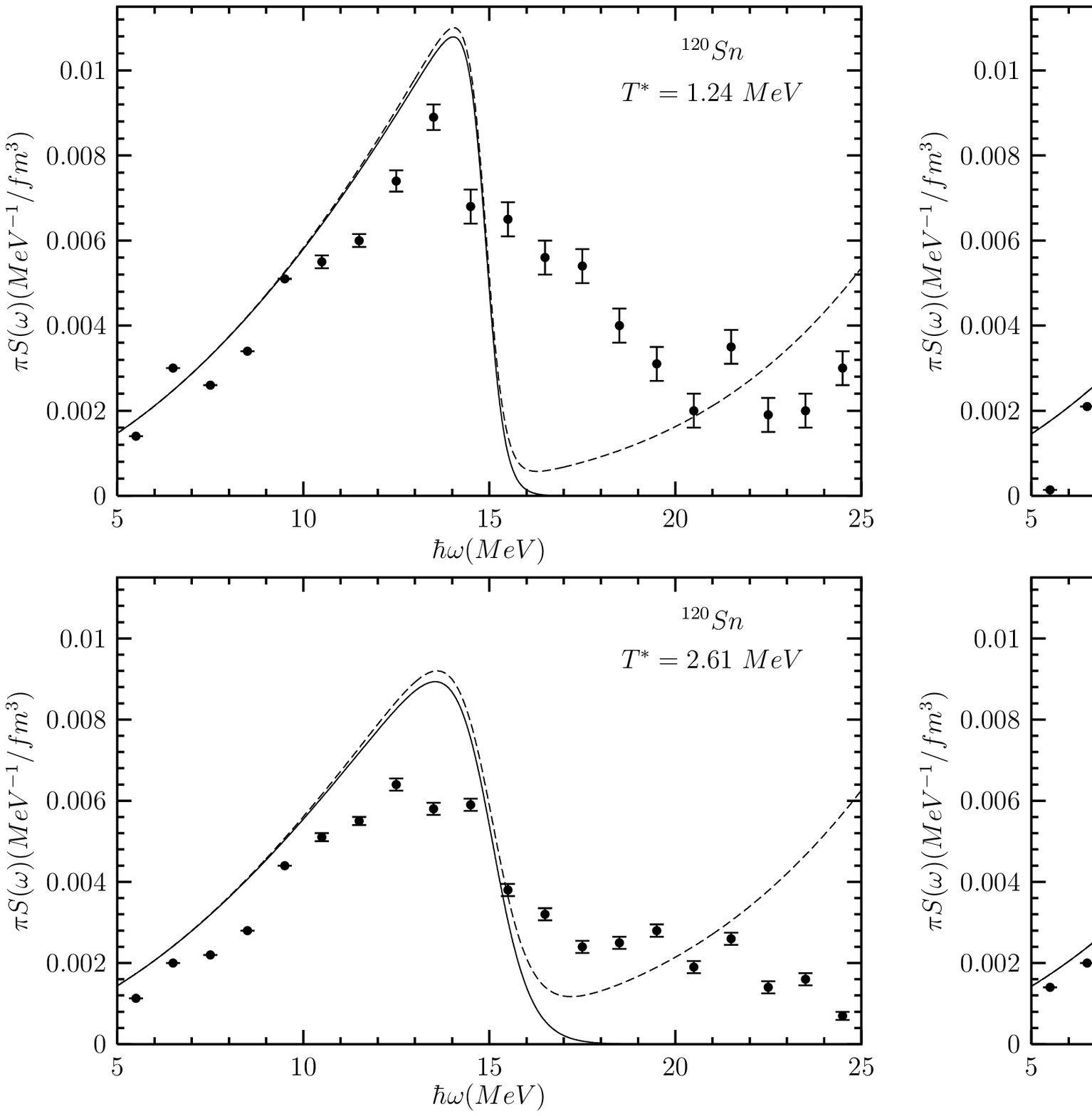,width=13cm,height=22cm,
angle=0}\vspace*{-7cm} \caption{The GDR strength function of
$^{120}$Sn obtained using Steinwedel-Jensen model. Solid and
dashed lines show the response function without and with the
collision term, respectively. The normalized data is taken from
[1].}
\end{figure}

\begin{figure}
\vspace*{0.0cm}\hspace{-3.8cm}
\epsfig{figure=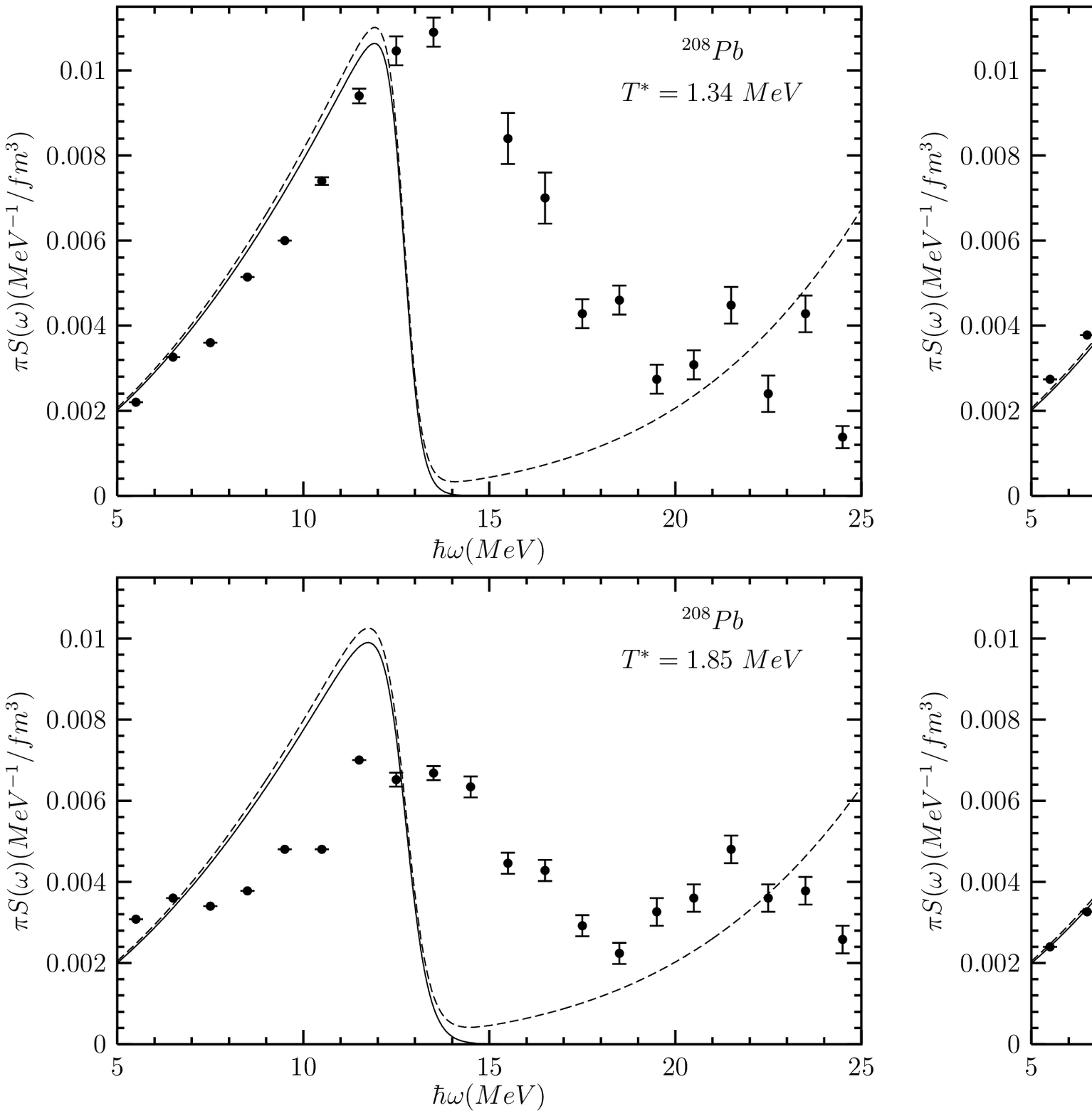,width=13cm,height=22cm,
angle=0}\vspace*{-7cm} \caption{The GDR strength function of
$^{208}$Pb obtained using Steinwedel-Jensen model. Solid and
dashed lines show the response function without and with the
collision term, respectively. The normalized data is taken from
[1].}
\end{figure}
\newpage

\begin{figure}
\vspace*{0.0cm}\hspace{-3.8cm}
\epsfig{figure=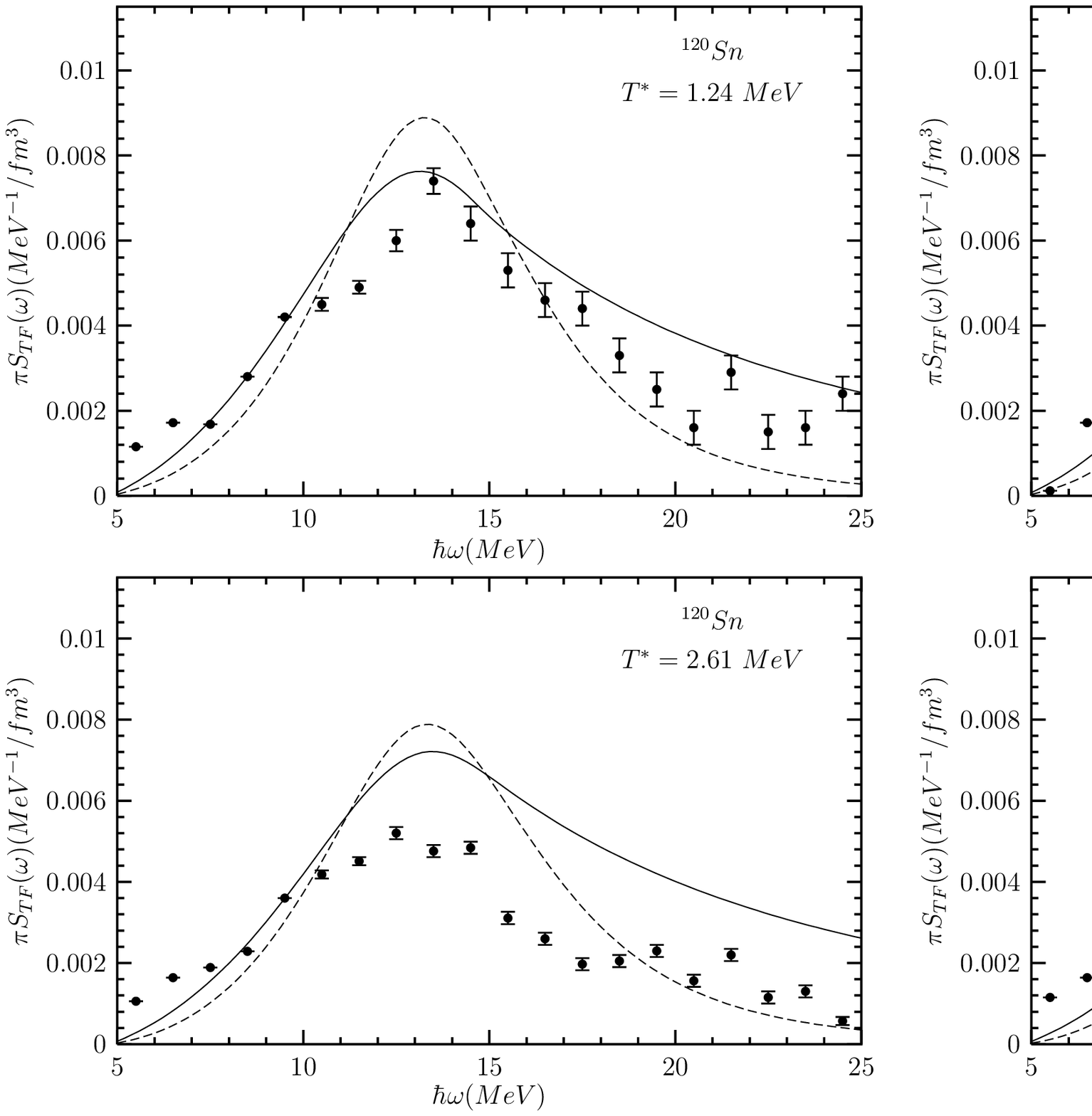,width=13cm,height=22cm,
angle=0}\vspace*{-7cm} \caption{The GDR strength function of
$^{120}$Sn calculated by Thomas-Fermi approximation. Solid and
dashed lines show the response function without and with the
collision term for the finite nuclear matter. The normalized data
is taken from [1].}
\end{figure}

\begin{figure}
\vspace*{0.0cm}\hspace{-3.8cm}
\epsfig{figure=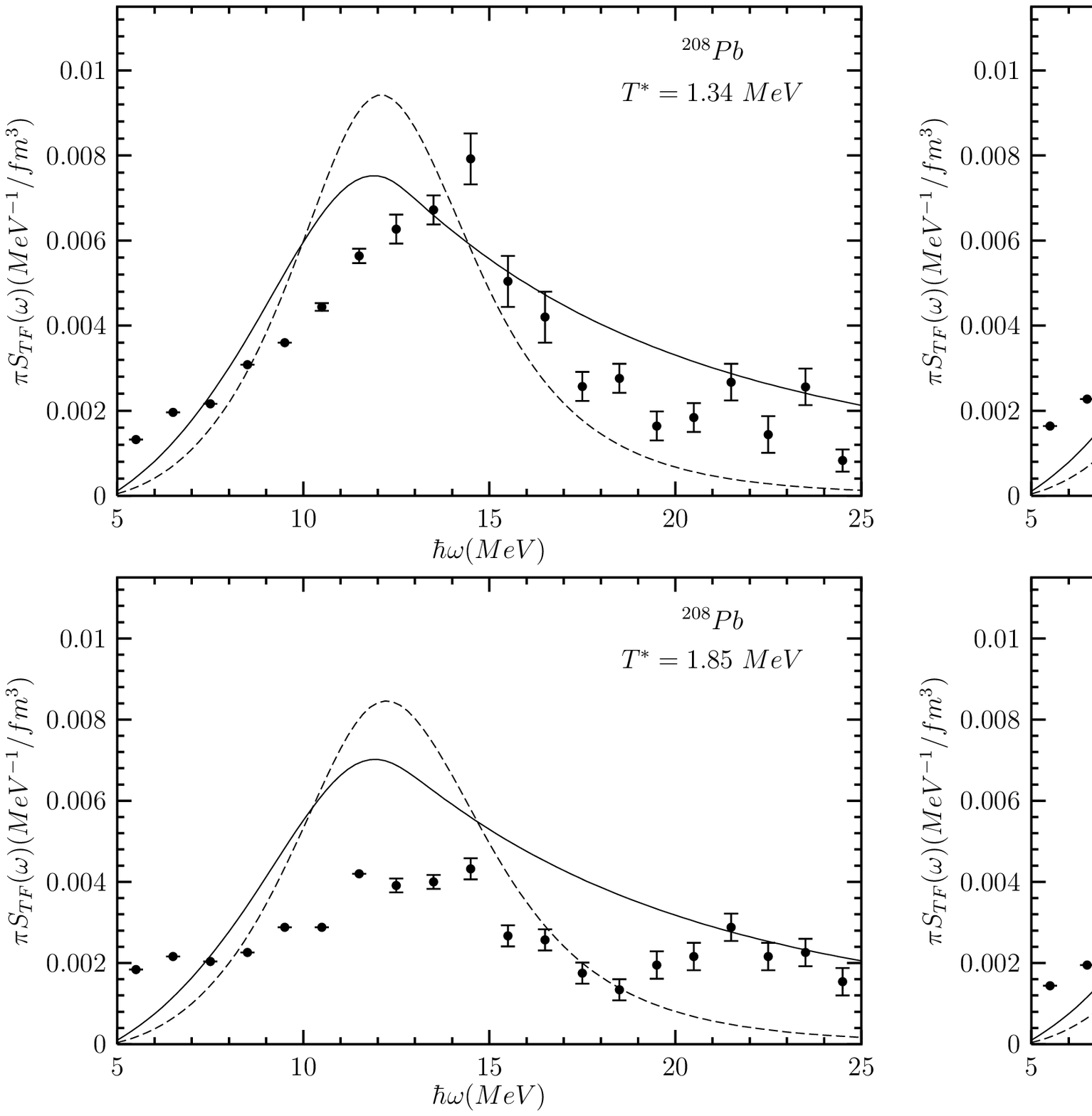,width=13cm,height=22cm,
angle=0}\vspace*{-7cm} \caption{The GDR strength function of
$^{208}$Pb calculated by Thomas-Fermi approximation. Solid and
dashed lines show the response function without and with the
collision term for the finite nuclear matter. The normalized data
is taken from [1].}
\end{figure}

\end{document}